\documentstyle[12pt,aps,epsf,floats]{revtex}
\vspace{2cm}
\begin{document}
\title{\ \\ \ \\ \ \\ \ \\ \ \\ \ \\LUTTINGER\ LIQUIDS COUPLED\ BY\ HOPPING.}
\author{Daniel Boies, Claude Bourbonnais and A.-M.S. Tremblay}
\address{D\'epartement de physique and Centre de recherche en physique du solide,}
\address{Universit\'e de Sherbrooke, Sherbrooke, Qc J1K 2R1, Canada.}
\maketitle

\thispagestyle{empty} \bigskip

{\small The stability of the Luttinger liquid to small transverse hopping
has been studied from several points of view. The renormalization group
approach in particular has been criticized because it does not take
explicitly into account the difference between spin and charge velocities
and because the interaction should be turned-on before the transverse
hopping if the stability of the Luttinger liquid is a non-perturbative
effect. An approach that answers both of these objections is explained here.
It shows that the Luttinger liquid is unstable to arbitrarily small
transverse hopping. The crossover temperatures below which either transverse
coherent band motion or long-range order start to develop can be finite even
when spin and charge velocities differ. Explicit scaling relations for these
one-particle and two-particle crossover temperatures are derived in terms of
transverse hopping, spin and charge velocities and anomalous exponents. The
renormalization group results are recovered as special cases when spin and
charge velocities are identical. The results compare well with recent
experiments presented at this conference. Magnetic field effects are alluded
to. }\bigskip

\section{Introduction.}

The traditional approach to solid-state Physics relies on the existence of
electronic elementary excitations that behave in many ways like
non-interacting electrons. They are the ``quasiparticles'' of Landau Fermi
liquid theory. Ever since Luttinger, it is known that in the case where
interacting electrons move in {\it one} dimension, the Fermi liquid
quasiparticles do not exist. Indeed, there are no simple fermionic
eigenstates of the Hamiltonian that adiabatically connect with the
non-interacting electrons, as in Fermi liquid theory. Instead, the excited
eigenstates of the Hamiltonian become spin and charge collective modes
(undamped paramagnons and zero-sound modes) each having their own linear
dispersion relation and corresponding velocity. The single-particle Green's
function looses its single-particle pole in favor of a branch cut. Since the
characteristic low-energy properties we have just described are displayed by
a large class of microscopic models, the corresponding phenomenology has
been given a generic name: Luttinger liquids.\cite{Haldane} The next section
will remind the reader of key properties of electrons interacting in
one-dimension.

In this conference proceedings, we study the problem of one-dimensional
chains of interacting electrons, (Luttinger liquids) coupled by a
perpendicular hopping amplitude $t_{\bot }$.This problem is of interest for
several reasons. It is a reasonable model of quasi-one-dimensional
conductors and possibly of coupled quantum wires that are discussed in this
conference. Also, this problem can possibly lead to insights into the
two-dimensional case that is of interest for high-temperature
superconductors. More specifically, we address the question: Is there a
finite critical value of $t_{\bot }$ below which the Luttinger liquid is
stable at zero temperature? P.W. Anderson \cite{Anders} has conjectured that
when spin and charge velocities are different, then the answer to this
question is yes. If there was such a critical value of $t_{\bot }$ below
which the Luttinger liquid would survive, in the jargon we would say that we
have ``confinement''. The problem that we are studying is also known as the
problem of ``dimensional crossover''. Indeed, when the perturbation $t_{\bot
}$ is zero, the system is clearly one-dimensional, while when the hopping
parallel and perpendicular to the chains are equal, the system is clearly
two-dimensional.

This question of the stability of the Luttinger liquid in the presence of $%
t_{\bot }$ has been addressed long ago in the context of
quasi-one-dimensional conductors \cite{Bourbon1,Brazo,Bourbon2}. There it
was shown that $t_{\bot }$ as well as pair-hopping correlations destabilize
the Luttinger liquid fixed point. A series of recent works on two coupled
chains \cite{Twochain} and on the many-chain problem \cite{Castel92} confirm
this view but, nevertheless, the issue remains controversial \cite{Anders3}.
In particular, going beyond the two chain case is a necessary requirement
for phase transitions or long-range quasi-particle coherence to occur. The
very few attempts to do so essentially deal with the situation where the
spin and charge velocities are equal \cite{Brazo,Bourbon2}. To check
Anderson's conjecture, we must thus consider an approach where spin and
charge velocities are different from the outset in the one-dimensional
problem. Furthermore, again following Anderson's suggestion, we must turn-on 
$t_{\bot }$ after the interaction, in such a way that we do start with fully
correlated Luttinger liquids that are then coupled by a transverse hopping
amplitude.

Since previous approaches do not satisfy the above two requirements, we
start from a new functional integral formulation that does allow a
systematic expansion in powers of $t_{\bot }$ using as the unperturbed
system the Luttinger liquid with both anomalous exponents {\it and}
differing spin and charge velocities. This allows us to generalize previous
results and to understand in a unified way various limiting cases obtained
before by several authors. We investigate both the single-particle spectral
weight as well as the induced two-particle correlations.

Our results have appeared in various forms before\cite{TheseBoies}\cite{prl}%
. We emphasize that, contrary to most previous investigations, we do not
work at zero temperature. Instead, we consider the experimentally relevant
situation where we start at finite temperature and study the stability of
the Luttinger liquid as we decrease temperature. It should be intuitively
clear that the Luttinger liquid is not destroyed by a perpendicular hopping
when the temperature is much larger than the hopping integral $t_{\bot }\ll T
$ ($k_B=1$) while remaining much smaller than the degeneracy temperature $%
T\ll E_F$. Indeed, in this limit the energy uncertainty brought about by
thermal fluctuations does not allow one to detect the curvature of the Fermi
surface that differentiates the quasi-one-dimensional case from the purely
one-dimensional case. This is illustrated schematically in the figure below,
drawn in the Brillouin zone of one-dimensional chains lying in a plane. On
the left, we see the case where the spread in wave-vector caused by the
thermal uncertainty $\Delta \epsilon _{{\bf k}}=\Delta \left( -2t_{\bot
}\cos k_{\bot }a_{\bot }\right) \sim T$ corresponds to a wave vector
uncertainty $\Delta k_{\bot }$ that is as large as the extent of the
Brillouin zone in the direction perpendicular to the chains. Using the
uncertainty principle, this means that the particles are essentially
localized along the chains $1/\Delta k_{\bot }\sim a_{\bot }$, $\left( \hbar
=1\right) $. The figure to the right shows the case where the temperature is
low enough that the curvature of the Fermi surface becomes relevant. In the
latter case, the nesting wave vector is the higher-dimensional one, contrary
to the high-temperature case on the left where the nesting wave vector is
one-dimensional.

\begin{figure}
\centerline{\epsfxsize 4cm \epsffile{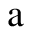}}
\caption{The solid lines represent the Fermi surface of 
non-interacting one-dimensional electron chains 
coupled by a small perpendicular hopping. (From p.315 of Ref.\protect\cite{LesHouches})
 The dashed line is the purely one-dimensional
Fermi surface.  
The dotted lines represent a typical uncertainty in wave vector caused 
by thermal fluctuations. Nesting wave vectors have arrows. The nesting is one-dimensional
on the left-hand side figure because $T>>t_\protect\bot$. In this regime, the behavior is that of
a Luttinger liquid. On the right-hand side figure, the nesting vector 
is two-dimensional because $T<<t_\bot$ .} 
\label{autonum}

\end{figure}

\begin{figure}
\centerline{\epsfxsize 7cm \epsffile{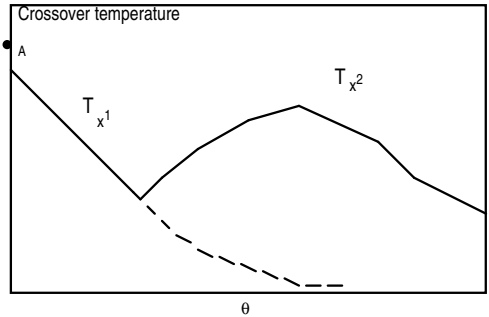}}
\caption{One-particle ($T_{x^1}$) and two-particle($T_{x^2}$) crossover
temperatures shown qualitatively as a
function of interaction strength schematically represented by $\theta$. The solid line becoming dotted
indicates the one-particle deconfinement Luttinger liquid (LL)$\rightarrow$Fermi Liquid (FL)
for the exact infinite-range transverse hopping model.
Point A gives the non-interacting value of $T_{x^1}$ ($\sim t_{\bot}/\pi$)
when $v_{\rho}=v_{\sigma}=v_F$.} 
\label{Diagramme}

\end{figure}

Our main conclusions can be summarized as follows. For a given bare $t_{\bot
}$ and temperatures much lower than the Fermi energy $T\ll E_F$, the typical
crossover diagram that we find is sketched in Fig.\ref{Diagramme} as a
function of temperature and of the anomalous exponent $\theta ,$ the latter
being taken as a measure of the interaction strength. As temperature is
lowered, two types of crossovers can occur. For weak enough interaction,
transverse one-particle coherent motion starts to develop, indicating that
the crossover at the deconfinement temperature $T_{x^1}$ is from Luttinger
liquid to Fermi liquid. This interpretation of $T_{x^1}$ is further
reinforced by the infinite-range hopping model\cite{These}\cite{prl} where
this result can be shown exactly. The one particle-crossover $T_{x^1}$ was
studied before to two-loop order by Prigodin and Firsov\cite{Bourbon1} and
to infinite order by Bourbonnais\cite{Bourbon1} and Schulz\cite{spinchar}.
Moving on to strong interactions, virtual pair hopping becomes the
dominating process which eventually leads to long-range ordering below the
two-particle dimensional crossover temperature $T_{x^2}$, even if there is
confinement at the one-particle level ($T_{x^2}\negthinspace >\negthinspace %
T_{x^1}$). At temperatures such that $0<$$T\negthinspace <T_{x^1}\left(
T_{x^2}\right) $, a true phase transition can occur in more than two
dimensions. It is important to notice that as long as there is no gap in the
two-particle excitations, the crossover temperature $T_{x^2}$ increases with
interaction strength, as shown in the figure. But when a gap appears, then
the crossover temperature $T_{x^2}$ starts to decrease with interaction
strength. This regime was studied by Brazovskii and Yakovenko\cite{Brazo}.
The reason for the decrease with interaction strength can be understood by
analogy with the Hubbard model at half-filling. In strong coupling, it maps
onto the Heisenberg model but with an exchange constant $J$ that decreases
with interaction strength as $J\sim t^2/U$.

It is clear that inasmuch as the crossover lines decrease to lower
temperature with decreasing $t_{\bot }$ but without disappearing, then our
results contradict those of Anderson. However, it is correct to say that it
is possible to go from a Luttinger liquid to long-range order without ever
becoming a Fermi liquid. We stress, furthermore, that our approach is
strictly valid only above the crossover lines. While we can only conjecture,
as above, what happens at temperatures below the crossover lines, we can, on
the other hand, clearly state that the behavior is not that of a Luttinger
liquid.

In the rest of this paper, we first recall some properties of the Luttinger
liquid, in part to set the notation in the context of other presentations at
this conference. Then we introduce our formalism and apply it to dimensional
crossover at the one- and two-particle level ( $T_{x^1},T_{x^2}$). In the
one-particle case, we briefly comment on the effect of a magnetic field.

\section{Some properties of the Luttinger liquid.}

\subsection{One-particle properties}

As a simple example of a Luttinger liquid, consider a Hamiltonian with
electrons moving to the right or to the left with velocities $\pm v_F$. The
kinetic energy is linear in the difference between the wave vector and the
Fermi wave vector, and we assume that right-moving electrons interact only
with right-moving electrons of opposite spin through a momentum independent
two-body interaction $g_4$, and similarly for left-moving electrons. Then at
zero temperature, the one-body Green's function for right-moving electrons
has the following asymptotic form 
\begin{equation}
\label{GreenG4}G_{+}\left( x,t\right) \sim \frac{e^{ik_Fx}}{\sqrt{x-v_\sigma
t}\sqrt{x-v_\rho t}}
\end{equation}
where the $+(-)$ index to $G$ refers to right (left) moving electrons. The
spin velocity $v_\sigma $ and the charge velocity $v_\rho $ are given by 
\begin{equation}
v_\sigma =v_F\left( 1-\frac{g_4}{2\pi v_F}\right) \quad ;\quad v_\rho
=\left( 1+\frac{g_4}{2\pi v_F}\right) 
\end{equation}
The spectral weight, or probability that an electron of momentum ${\bf k}$
has an energy $\omega $ is given, for right-moving electrons, by 
\begin{equation}
A\left( {\bf k},\omega \right) =-\pi ^{-1} {\rm Im} G_{+}^R\left( {\bf 
k},\omega
\right) =\frac{-1}\pi {\rm Im} \left[ \frac 1{\sqrt{\omega +i\eta -v_\sigma 
k}} \frac 1{\sqrt{\omega +i\eta -v_\rho k}}\right] 
\end{equation}
Here, $k$ is measured with respect to the Fermi point. The spectral weight
is schematically represented in the following figure. It is clearly
qualitatively different from a Fermi liquid where we would see a peak
representing a quasiparticle standing on a broad incoherent background.

\begin{figure} \centerline{\epsfxsize 6cm \epsffile{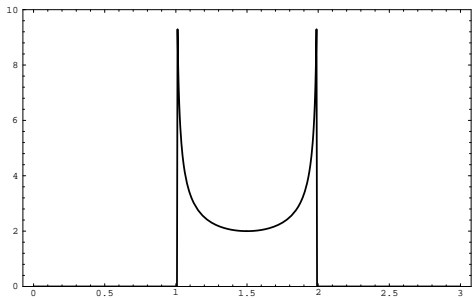}} 
\caption{Schematic representation of the spectral weight in the $g_4$ model} 
\label{SpectralG4Fig} \end{figure}

Note that the single-particle Green's function Eq.(\ref{GreenG4}) {\it in
real space }factors into a part that comes from collective spin excitations
and a part that comes from collective charge excitations. This qualitative
feature is a characteristic of Luttinger liquids. This is what is meant by
spin-charge separation.\cite{spinchar} In more general models, one will have
power laws with interaction-dependent exponents instead of the square roots
appearing in the simple $g_4$ model. The spin velocity and the exponent
associated with spin will appear in one factor, while the charge velocity
and the exponent associated with charge will appear in another factor. In
fact, one of the factors, spin or charge, can even be gaped while the other
is still described by a gapless spectrum.

The use of powerful techniques, such as the renormalization group, the
Bethe ansatz, bosonization and conformal field theory techniques have
allowed a quite thorough understanding of physical properties of Luttinger
liquids.
properties, such as the asymptotic low-energy behavior of the {\it
g}-ology model, have been known for a long
time\cite{geo}\cite{lutt}\cite{dzyalo}. By contrast, the calculation of
asymptotic properties of correlation functions of the Hubbard model with
and without magnetic field is relatively recent\cite{frahm}\cite
{frahmdeux}. 

In the absence of either a spin or a charge gap, the most general form that
the finite temperature one-particle Green's function can take for
right-moving electrons in a Luttinger liquid is $\cite{geo}\cite{dzyalo}\cite
{emery}\cite{frahm}$ 
\begin{equation}
\label{luttpropa}G_{+}^{\left( 1\right) }\left( x,\tau \right) =\frac{%
e^{ik_F^0x}}{2\pi i\Lambda ^\theta }\prod_{\nu =\rho ,\sigma }S\left( z_\nu
\right) ^{-1/2-\theta _\nu /2}S\left( z_\nu ^{*}\right) ^{-\theta _\nu /2}
\end{equation}
where $S\left( z_\nu \right) \equiv \xi _\nu \sinh \left( z_\nu /\xi _\nu
\right) $ and where naturally appears the complex coordinate $z_\nu
=x+iv_\nu \tau ,$ with $\tau $ the Matsubara time. The other symbols are the
subscript $p=\pm $ to $G_p$ for the branch index and $\Lambda $ for the
ultraviolet cutoff in wave number. The non-universal exponents $\theta _\rho
>0$ and $\theta _\sigma >0$ are associated respectively with charge and
spin, while the exponent without subscript, that we call the propagator's
anomalous exponent $\theta $, is simply $\theta =\theta _\rho +\theta
_\sigma $. As before, $v_\rho $ et $v_\sigma $ are respectively the charge
and spin velocities, while the associated thermal de Broglie correlation
lengths are $\xi _{\rho ,\sigma }=v_{\rho ,\sigma }/\pi T$, $\left( \hbar
=1\right) .$ For values of space and imaginary time larger than these
correlation lengths, the propagator falls-off exponentially, while for
values much smaller than these lengths, it has a power law behavior, a
phenomenology consistent with that of propagators close to a critical point.
This critical point is at zero temperature since this is where the coherence
lengths become infinite (Since the ratio of these lengths is a constant,
there is only one of them that is relevant.) The existence of this
zero-temperature critical point allows one to make full use of the
renormalization group approach, as shown in Ref.\cite{Bourbon2}.

Unfortunately, many different notations exist in this field. For the
convenience of the reader, we mention that often in this conference the case
of no magnetic anisotropy $\theta _\sigma =0$ was considered. Furthermore,
the notation 
\begin{equation}
\theta _\rho =\frac 14\left( K_\rho +K_\rho ^{-1}-2\right) 
\end{equation}
was often used. Without attempting to be exhaustive, we give the following
rosetta stone\cite{TheseBoies} (Table I) that may help the reader move
through the literature. The first line is the notation used in the present
work and in earlier ones by Bourbonnais.\cite{Bourbon2}\cite{These}

\begin{table}[t]
\protect\label{UneParticule}
\begin{tabular}{cccc}
$\nu = ( \sigma , \rho )$
exponent & Anomalous exponent & $\nu = ( \sigma , \rho )$ velocities &
References \\ 
\hline 
$\theta _{\nu }$ & $\theta$ & $v_{\nu}$ & this work\\
$\eta _{\nu}$ & $\eta$ & $u_{\nu}$ & \cite{Castel92} ,\cite{casteldeux} \\
$2\alpha _{\nu}$ & $2\alpha$ & $u_{\nu}$ & \cite{geo}\\ 
$2\gamma
_{\nu}=\left( K_{\nu}+K_{\nu}^{-1}-2\right) /4$ & $2\gamma$ & $u_{\nu}$ &
\cite{Voit} \\ 
$\theta_{\rho}=\nu_1-1/2,\ \theta_{\sigma}=0$ & $\nu_1
-1/2$ & $v_{c,s}$ & \cite{frahm},\cite{frahmdeux}\\
$\eta_{F\nu}-1/2=\left( \gamma_{\nu}+\gamma_{\nu}^{-1}-2\right) /4$ &
$\eta_{F}-1$ & $v_{\rho,s}$ & \cite{brazoun} ,\cite{brazodeux}\\
\end{tabular} 
\caption{Connection between different notations for
one-particle properties} 
\end{table}

\subsection{Two-particle properties}

To study our problem, we will also need the asymptotic behavior of
two-particle, or four point connected functions of the Luttinger liquid. By
definition, omitting temporarily branch and spin indices, the four-point
connected function is defined by 
\begin{equation}
G_c^{\left( 2\right) }\left( \left\{ z\right\} \right) =G^{\left( 2\right)
}\left( \left\{ z\right\} \right) +G^{\left( 1\right) }\left( z_1-z_4\right)
G^{\left( 1\right) }\left( z_2-z_3\right) -G^{\left( 1\right) }\left(
z_1-z_3\right) G^{\left( 1\right) }\left( z_2-z_4\right) 
\end{equation}
where $\left\{ z\right\} =\left( z_1,z_2,z_3,z_4\right) $ and, restoring
indices, 
\begin{equation}
G_{\left\{ p,s\right\} }^{\left( 2\right) }\left( \left\{ z\right\} \right)
=\left\langle T_\tau a_{p_1}^{s_1}\left( z_1\right) a_{p_2}^{s_2}\left(
z_2\right) a_{p_3}^{s_3\dagger }\left( z_3\right) a_{p_4}^{s_4\dagger
}\left( z_4\right) \right\rangle _{1D}
\end{equation}

There are essentially three regimes with different asymptotic behavior. One
weak-coupling regime, and two strong-coupling regimes. One of the strong
coupling regime is gapless, while the other one has a gap. In the
weak-coupling regime, the four-point function essentially scales as the
product of two one-particle Green's functions. Hence $\theta _\rho $ and $%
\theta _\sigma $ remain the only relevant exponents in this regime. By
contrast, in the gapless strong-coupling regime, the four-point function
scales as a product of composite bosonic operators. Hence, two new exponents
appear, $\gamma _\rho $ and $\gamma _\sigma $. At even stronger coupling, a
gap may appear in either the spin or the charge degrees of freedom. It is
also important to realize that higher-order connected functions, $%
G_c^{\left( n\right) }$ for $n\geq 3,$ do not involve new exponents. Let us
give the results for the two strong-coupling regimes.

\subsubsection{Gapless strong-coupling regime.}

The non-vanishing two-particle Green's
functions of a Luttinger liquid are of the form $G_{p,-p,-p,p}^{\left(
2\right) }\left( z_1 s ,z_2 s ^{\prime },z_3 s^{\prime
},z_4 s
\right) $. Within logarithmic factors, their asymptotic form obtained from
bosonization is\cite {brazoun}\cite{brazodeux}%
$$
G_{s,s'}^{\left( 2\right) }\left( \left\{ z\right\} \right) \sim
\Lambda ^{-2\theta }\prod_{\nu=\rho ,\sigma } \mid S\left( z_{1\nu
}-z_{4\nu }\right)S\left( z_{2\nu }-z_{3\nu }\right)\mid ^{-1/2-\theta
_\nu }
$$
\begin{equation}
\label{biggreen}\times  \mid S\left( z_{1\nu }-z_{3\nu }\right)
S\left( z_{2\nu }-z_{4\nu }\right)\mid^{-\delta_\nu(s,s')}
\mid S\left( z_{1\nu }-z_{2\nu }\right) S\left( z_{3\nu
}-z_{4\nu }\right)\mid^{-\bar{\delta}_\nu(s,s')}
\end{equation}
where $\delta_\rho(s,s')= \theta_\rho- \gamma_\rho/2 + 1/2 =
-\bar{\delta}_\rho(s,s')$ (independent of $s, s'$) for charge degrees of
freedom, while for spin degrees of freedom, $\delta_\sigma(s=s')=
\theta_\sigma- \gamma_\sigma/2 + 1/ 2=-\bar{\delta}_\sigma(s= s')$, and
$\delta_\sigma(s\ne s')= \theta_\sigma- \gamma_\sigma^{-1}/ 2 + 1/2=
-\bar{\delta}_\sigma(s\ne s')$. The exponents $\gamma_{\rho}$ and
$\gamma_{\sigma}$ are functions of the microscopic interactions $g_{i\Vert
}, g_{i\bot }$. When there is no spin anisotropy, namely when $g_{i\Vert
}=g_{i\bot }$, the spin-related exponent is $\gamma _\sigma =1$ while
$\gamma_\rho$ is non-universal\cite{geo}\cite{emery}. The relation to
one-particle exponents is,\cite {emery}

\begin{equation}
\theta _\nu =\frac 14\left( \gamma _\nu +\gamma _\nu ^{-1}-2\right) \text{.}
\end{equation}
The gapless strong-coupling regime is defined as follows. Let us first
consider the (repulsive) particle-hole channel. In strong coupling, the
correlation function decays more rapidly than the inverse square of the
relative particle-hole distances $\left| z_{1\nu }-z_{3\nu }\right| $ et $%
\left| z_{2\nu }-z_{4\nu }\right| $ so that the dominant asymptotic behavior
is given by the relative distance between the two particle-hole pairs $%
z=z_1-z_2$. This situation is realized when $\delta _\rho \left( s,s^{\prime
}\right) +\delta _\sigma (s,s^{\prime })>1$. For charge-density-wave (CDW)
correlations, $s=s^{\prime }$, and one has $2\theta -\gamma _\rho -\gamma
_\sigma >0$ whereas for spin-density-wave (SDW) correlations, $s\ne
s^{\prime }$ and the strong coupling condition reads $2\theta -\gamma _\rho
-\gamma _\sigma ^{-1}>0$. The strong coupling condition for electron-hole
correlations amounts to put $\left| z_{1\nu }-z_{3\nu }\right| \sim d_{\Vert
}$ and $\left| z_{2\nu }-z_{4\nu }\right| \sim d_{\Vert }$ in the regime $%
\xi _\nu \gg d_{\Vert }\sim \Lambda ^{-1}$, so that we have $\left| S\left(
d_{\Vert }+id_{\Vert }\right) \right| \sim d_{\Vert }$. Using this result
and the definitions 
\begin{equation}
\label{gammamu}\gamma _\mu =2-\gamma _{\mu ,\rho }-\gamma _{\mu ,\sigma }
\end{equation}
$\mu =0$ for CDW and $\mu \ne 0$ for SDW with $\gamma _{0,\rho }=\gamma
_\rho $, $\gamma _{0,\sigma }=\gamma _\sigma $ and $\gamma _{\mu \ne 0,\rho
}=\gamma _\rho $, $\gamma _{\mu \ne 0,\sigma }=\gamma _\sigma ^{-1}$, we can
write 
\begin{equation}
\label{allo}G_\mu ^{\left( 2\right) }\left( z_1-z_2\right) \sim \Lambda
^{\gamma _\mu }\prod_{\nu =\rho ,\sigma }\left| S\left( z_{1\nu }-z_{2\nu
}\right) \right| ^{-\gamma _{\mu ,\nu }},
\end{equation}
($\Lambda ^{\gamma _\mu }\sim $ $\Lambda ^{-2\theta }d_{\Vert }^{-2\theta
-\gamma _\mu })$. Note that this asymptotic behavior is the same as that of
the response function $R_\mu \left( z\right) $ calculated in the same
channel $\mu $\cite{emery}\cite{These} 
\begin{equation}
\label{repos}R_\mu \left( z\right) \sim \Lambda ^{\gamma _\mu }\cos
2k_F^0x\prod_{\nu =\rho ,\sigma }\left[ f_\nu \left( z_\nu ,z_\nu
^{*}\right) S\left( z_\nu \right) S\left( z_\nu ^{*}\right) \right]
^{-\gamma _{\mu ,\nu }/2}
\end{equation}
where $f_\nu \left( z_\nu ,z_\nu ^{*}\right) =2\left| z_\nu \right| ^2\left[
z_\nu ^2+z_\nu ^{*2}\right] ^{-1}$ is a dimensionless function and where $%
z=z_1-z_2$ is the relative distance between pairs. The Fourier transform of
(11) is well known to give rise to a power law singularity $R_\mu
(2k_F^0,T)\sim T^{-\gamma _\mu }$ in temperature. With the above
definitions, the strong-coupling condition in the electron-hole channel can
be written in the form 
\begin{equation}
\label{Fort}2-2\theta -\gamma _\mu <0
\end{equation}

Similarly, strong attractive coupling in the electron-electron channel is
achieved when $\bar \delta _\rho (s,s^{\prime })+\bar \delta _\sigma
(s,s^{\prime })>1$, which for triplet superconductivity (TS), $s=s^{\prime
}$ , corresponds to the condition $2\theta -\gamma _\rho ^{-1}-\gamma
_\sigma ^{-1}>0$, while for singlet superconductivity (SS), $s\ne
s^{\prime }$, it corresponds to the condition $2\theta -\gamma _\rho
^{-1}-\gamma _\sigma >0$
superconducting correlation and response functions [Eqs.
(\ref{allo})(\ref{repos})] with $\gamma _{\mu =0,\rho }=\gamma _\rho
^{-1},\gamma _{\mu =0,\sigma }=\gamma _\sigma $ for SS($\mu =0$) and
$\gamma _{\mu \ne 0,\rho }=\gamma _\rho ^{-1},\gamma _{\mu \ne 0,\sigma
}=\gamma _\sigma ^{-1}$ for TS($\mu \ne 0$). With these definitions of
$\gamma _\mu $ in the particle-particle channel, the power law singularity
in temperature of the response functions is again $R_\mu (T)\sim
T^{-\gamma _\mu }$ with the same functional form for both
Eq.(\ref{gammamu}) $\gamma _\mu =2-\gamma _{\mu ,\rho }-\gamma _{\mu
,\sigma }$ and the strong-coupling condition Eq.(\ref{Fort}). 

\subsubsection{Strong-coupling regime with a gap.}

A gap can open up for example at half-filling when umklapp scattering $g_3$
becomes relevant. Then, the charge excitations become gapped when\cite{emery}
$\left| g_3\right| +2g_2-g_1>0$. The asymptotic form of the two-body
correlation function in the presence of such a gap $\Delta _\rho $ is known%
\cite{gutklem}. We have been able to extract the result when spin and charge
velocities differ. Let $l_\rho \sim v_\rho /\Delta _\rho $ be the size of a
particle-hole pair. Then, when the distance between the particle and the
hole forming a pair satisfies $\left| z_{1\rho }-z_{3\rho }\right| ,\left|
z_{2\rho }-z_{4\rho }\right| \ll l_\rho $, while the interpair distance is
such that $\left| z_{1\rho }-z_{2\rho }\right| \gg l_\rho $ then the
asymptotic form is 
\begin{equation}
\label{gaprho}G^{\left( 2\right) }\left( \left\{ z\right\} \right) \sim
\Lambda ^{-2\theta }\frac{e^{-\left| z_{1\rho }-z_{3\rho }\right| /l_\rho
-\left| z_{2\rho }-z_{4\rho }\right| /l_\rho }}{l_\rho ^{1+2\theta _\rho
}\left| S\left( z_{1\sigma }-z_{2\sigma }\right) \right| ^{1+2\theta _\sigma
}}\text{.}
\end{equation}
Clearly, the thermal correlation length $\xi _\rho $ is replaced by the
characteristic size of the pair $l_\rho $. Note that one can find an
expression analogous to the last one, Eq.(\ref{gaprho}), when there is a gap 
$\Delta _\sigma $ in the spin channel, as may happen in the presence of
attractive backscattering $g_1<0$.

As in the one-particle case, we present the (non-exhaustive) table II that
should help the reader through the literature. The first line refers to the
notation of Refs.\cite{Bourbon2}\cite{These} and of the present work.

\begin{table}[t]
 \protect\label{DeuxParticules}
 \begin{tabular}{ccc}
 $\nu = (\sigma , \rho )$ exponent & Susceptibility exponent & References \\   \hline

$\gamma _{\mu,\nu }$ & $\gamma_\mu=2-\gamma _{\mu,\rho }-\gamma 
_{\mu,\sigma }$ & this work\\
$\theta _{(c, s)}^{\mu}=\gamma_{\mu, (\rho ,\sigma)}$ & - & \cite{emery}\\
$\beta _{(c, s)}^{\mu}=\gamma_{\mu, (\rho ,\sigma)}$ & - & \cite{geo} \\
$\theta=4\gamma_{\mu,\rho}, \gamma_{\mu,\sigma}=1$ & - & \cite{frahm}
,\cite{frahmdeux}
 \end{tabular}
 \caption{Connection between different notations for two-particle properties}
 \end{table}

Note that in Refs.\cite{geo} and \cite{emery}, the index $\mu $ means that $%
\beta _c^{0,1}=\beta _c$ and $\beta _c^{2,3}=\beta _c{}^{-1}$ while $\beta
_s^{0,2}=\beta _s$ and $\beta _s^{1,3}=\beta _s{}^{-1}$.

\section{Formalism for dimensional crossover.}

Let us start with the full partition function for a set of $N_{\bot }$ fully
interacting chains written in the interaction representation where the
unperturbed Hamiltonian is the complete interacting one-dimensional
Hamiltonian 
\begin{equation}
{Z={\rm Tr}\{e^{-\beta \left( \sum_i{{\cal H}_i^{1D}}+\sum_{ij}{{{\cal H}%
_{\bot ij}}}\right) }\}=Z_{1D}\langle {T_\tau e^{-\int_0^\beta d\tau {%
\sum_{ij}}{\cal H}_{\bot ij}\left( \tau \right) }}\rangle _{1D}}
\end{equation}
The indices $i,j$ run over all $N_{\bot }$ chains, ${\cal H}_i^{1D}$ is the
purely $1D$ Hamiltonian describing the interacting electrons along chain $i$
while the interchain hopping part ${{{{\cal H}_{\bot ij}}}}$ stands as the
perturbation. The above thermodynamic average $\langle {\ldots }\rangle _{1D}
$ and~partition function~$Z_{1D}$ only involve the pure $1D$ Hamiltonian.
The hopping Hamiltonian is given by ${{\cal H}_{\bot ij}=-\int
dx\sum_{p,s}t_{\bot ij}{a_{p,i}^{s\dagger }}(x)a_{p,i}^s(x)}$ where $x$ is
the continuous coordinate along the chains while $p=\pm $ denote right and
left-going electrons as before. By analogy with the problem of propagation
of correlations in critical phenomena, the propagation of one-particle
transverse coherence is studied through an effective field theory which is
generated by a Hubbard-Stratonovich transformation for Grassmann variables.
This allows the partition function $Z$ to be expressed as a functional
integral over a Grassmann $\psi ^{\left( *\right) }$ field\widetext
\begin{equation}
\label{un}Z=Z_{1D}\negthinspace \int \negthinspace \negthinspace
\negthinspace \int {\cal D}\psi ^{*}{\cal D}\psi \,e^{\;-\int d(1)%
\negthinspace \sum\limits_{ij}\psi _i^{*}(1)t_{\bot ij}^{-1}\psi _j(1)}{%
\langle \;e^{\;\int d(1)\negthinspace \sum\limits_i\left( a_i^{\dagger
}(1)\psi _i(1)+\psi _i^{*}(1)a_i(1)\right) }\;\rangle _{1D}}
\end{equation}
with the notations $\int d(n)\psi _i(n)\equiv \int
dz_n\sum\nolimits_{p_n,s_n}\psi _{p_n,i}^{s_n}(z_n)$ and $\int dz_n\equiv
\int dx_n\int_0^\beta d\tau _n$. The logarithm of the thermodynamic average
in (\ref{un}) is readily recognized as the generating function for the exact 
$1D$ connected many-body Green's functions $G_c^{(n)}$. Changing variables
from $\psi _i$ to $\sum_jt_{\bot ij}^{1/2}\psi _j$ in the functional
integral (\ref{un}) allows us to obtain a field theory whose multipoint
interactions involve successively higher powers of $t_{\bot }.$ We find 
\begin{equation}
\label{deux}{Z}=Z_{1D}\int \negthinspace \negthinspace \negthinspace \int 
{\cal D}\psi ^{*}{\cal D}\psi \,e^{\,{\cal F}[\psi ^{*},\psi ]}
\end{equation}
where the Grassmannian Landau-Ginzburg-Wilson functional ${\cal F}={\cal F}%
_0+\sum\nolimits_{n\ge 2}{\cal F}^{(n)}$ involves a quadratic part and a sum
of effective interactions to all orders in the $\psi $ field. To write a
specific form for ${\cal F}$, let us consider the case where chains are
lined up in a plane and let us Fourier transform in the direction transverse
to the chains. Then, the Gaussian part describing the free propagation of
the $\psi $ field takes the form \widetext
\begin{equation}
\label{trois}{\cal F}_0=-\int \negthinspace d(1)\negthinspace \int 
\negthinspace d(2)\negthinspace \sum\limits_{k_{\bot }}\psi _{k_{\bot
}}^{*}(1)\left( \openone -t_{\bot }(k_{\bot })G^{(1)}(1-2)\right) \psi
_{k_{\bot }}(2),
\end{equation}
where $G^{(1)}$ is the exact one-dimensional propagator. The interacting
part is found to be 
\begin{equation}
\label{quatre}{\cal F}^{(n)}={\frac 1{\left( n!\right) ^2}\frac 1{N_{\bot
}^{n-1}}}\sum\limits_{k_{\bot 1}\ldots k_{\bot 2n}}\negthinspace \int 
\negthinspace
d(1)\ldots \int \negthinspace d(2n)\psi _{k_{\bot 1}}^{*}(1)\ldots \psi
_{k_{\bot 2n}}(2n)\gamma _{\bot }\left( k_{\bot 1}\ldots k_{\bot 2n}\right)
G_c^{(n)}(1\ldots 2n)
\end{equation}
where 
\begin{equation}
\label{gamma}\gamma _{\bot }\left( k_{\bot 1}\ldots k_{\bot 2n}\right)
=\left\{ \prod_\alpha ^{2n}t_{\bot }\left( k_{\bot \alpha }\right) \right\}
^{1/2}\delta _{\Sigma _\alpha k_{\bot \alpha },0}\ \;
\end{equation}
in which $t_{\perp }(k_{\perp })$ is the Fourier transform of $t_{\perp ij}$%
. The Gaussian part gives the exact result in a few special cases: a) In the
non-interacting limit $(G_c^{(n\geq 2)}\rightarrow 0)$ for arbitrary $%
t_{\bot }$ b) For arbitrary interaction when $t_{\bot }=0$. c) In the limit
of infinite-range transverse hopping, whatever the value of $t_{\bot }$ and
of the interaction\cite{These}\cite{prl}. Therefore $1/N_{\bot }$ may be
used as a formal expansion parameter.

We will see that the crossover temperatures are those where the above
functional integral approach ceases to be valid. Dimensional considerations
like those presented below then lead to the following expansion parameters
in the three regimes previously identified: a) in weak-coupling $g/\pi
v_F\left[ t_{\bot }/E_F^\theta T^{1-\theta }\right] ^2$ with $g\ll \pi v_F$
the running coupling constant b) in the gapless strong (forward-scattering)
coupling regime , $\left[ t_{\bot }/E_F^{1-\gamma _\mu /2}T^{\gamma _\mu
/2}\right] ^2$ c) in the regime where there is, let us say, a charge $\Delta
_\rho $ gap, $\left[ t_{\bot }/E_F^\theta T^{1/2-\theta _\sigma }\Delta
_\rho ^{1/2-\theta _\rho }\right] ^2$.

\section{Dimensional crossover.}

As described in the introduction, the Luttinger liquid may become unstable
either at the one-particle or at the two-particle level. We describe these
two cases in turn. In the one-particle case, we treat a simple example all
the way to zero temperature, to suggest how a quasi-particle peak may
develop. We also mention the results for the one-particle crossover in the
case where a magnetic field is applied. For the explicit form of all
dimensionless functions $F_l$ that appear in the following results for the
crossover temperatures see Ref.\cite{TheseBoies}.

\subsection{Single-particle dimensional crossover.}

\subsubsection{Zero magnetic field}

An instability of the Luttinger liquid at zero temperature and thus the
possibility of a Fermi liquid fixed point is already present in the free
theory of the $\psi $ field described by ${\cal F}_0$. At this level of
approximation, the $2D$ one-particle Green's function in Fourier-Matsubara
space, say for right-going electrons, is given by 
\begin{equation}
\label{six}{\cal G}^{2D}({\bf k},\omega _n)={\frac{{G^{(1)}(k,\omega _n)}}{%
1-t_{\bot }(k_{\bot })G^{(1)}(k,\omega _n)}}, 
\end{equation}
where $\omega _n\negthinspace =\negthinspace (2n+1)\pi T$, while ${\bf k}%
=(k,k_{\bot })$ with $k$ measured with respect to the $1D$ Fermi point $%
k_F^0 $. For the propagator $G^{(1)}$ describing the Luttinger liquid in
space and imaginary time we can use our previous result Eq.(\ref{luttpropa}%
). At high temperature, $t_{\bot }(k_{\bot })G^{(1)}(k,\omega _n)\ll 1$ so
that we have Luttinger liquid behavior. The one-particle propagator will
definitely be different from that of a Luttinger liquid when the temperature
is sufficiently low that $t_{\bot }(k_{\bot })G^{(1)}(k,\omega _n)\sim 1$.
Although we cannot strictly say towards what state the crossover will be as
we decrease temperature, the following simple example suggests how a
quasiparticle pole may appear, even when we have differing spin and charge
velocities.

\begin{description}
\item[A simple $T=0$ example.]  Let us consider the Luttinger liquid Eq.(\ref
{luttpropa}) in the special case presented to introduce the Luttinger liquid
Eq.(\ref{GreenG4}). Then, $\theta _\rho \negthinspace =\negthinspace \theta
_\sigma \negthinspace
=\negthinspace 0$ but $v_\rho \negthinspace \neq \negthinspace
v_\sigma $. In this case, we saw that the $T\negthinspace =\negthinspace 0$
retarded $1D$ propagator $G^{(1)}(k,\omega )$ has two square root
singularities. At the Gaussian level, the corresponding spectral weight $%
A^{2D}=-\pi ^{-1}{\rm Im}{\cal G}^{2D}$ obtained from the imaginary part of
Eq.(\ref{six}) is given by 
\begin{equation}
\label{neuf}A^{2D}({\bf k},\omega )={\frac{A^{1D}(k,\omega )}{{1+\left( \pi
t_{\bot }(k_{\bot })A^{1D}(k,\omega )\right) ^2}}}+Z(t_{\bot })\delta
(\omega -\epsilon ({\bf k))}
\end{equation}
where $A^{1D}=-\pi ^{-1}{\rm Im} G^{\left( 1\right) }$ is the exact 
one-dimensional
spectral weight, $Z\left( t_{\bot }\right) $ is the quasi-particle residue
and $\epsilon \left( {\bf k}\right) $ the pole of (\ref{six}). For the case
at hand, the undamped quasi-particle spectrum given by the pole of Eq.(\ref
{six}) is $\epsilon \left( {\bf k}\right) =k\overline{v}$ $-{\rm sign}%
\left\{ t_{\bot }\left( k_{\bot }\right) \right\} \sqrt{\left( k\Delta
v\right) ^2+t_{\bot }^2\left( k_{\bot }\right) }$ where $\overline{v}%
\negthinspace =\negthinspace \left( v_\rho +v_\sigma \right) /2$ and $\Delta
v\negthinspace =\negthinspace \left( v_\rho -v_\sigma \right) /2$, while the
quasi-particle residue takes the form 
\begin{equation}
\label{dix}Z\left( t_{\bot }\right) =\frac{\left| t_{\bot }\left( k_{\bot
}\right) \right| }{\sqrt{\left( k\Delta v\right) ^2+t_{\bot }^2\left(
k_{\bot }\right) }}.
\end{equation}
The residue is readily seen to satisfy $Z\negthinspace \rightarrow 
\negthinspace 0$ when $t_{\bot }\negthinspace \rightarrow \negthinspace 0$
while $Z\negthinspace \rightarrow \negthinspace 1$ when $\Delta v%
\negthinspace \rightarrow \negthinspace 0$ as one expects for free
electrons. Note that the infinite lifetime of the quasi-particles in Eq.(\ref
{neuf}) should become finite when one goes beyond the Gaussian approximation
thus allowing the $\psi ^{\prime }s$ to interact. This should be true except
at the Fermi surface where the lifetime should remain infinite because of
the usual phase space arguments. One can also check that for wave-vectors
close to the new Fermi surface, given by $(v_\rho v_\sigma )^{1/2}k=t_{\bot
}(k_{\bot })$, the single-particle spectral weight has the following
frequency dependence: For nearest-neighbor hopping and $k_{\bot }%
\negthinspace <\negthinspace \pi /2$, a quasi-particle peak is first
encountered as the frequency is increased, followed by an incoherent
background which is a smoothed version (without square-root singularity) of
the original Luttinger liquid. In other words, remnants of spin-charge
separation are left at high energies. This is illustrated qualitatively in
the following figure which suggests that even if a Fermi liquid is recovered
at low temperature, it is strongly renormalized with many properties that
may be controlled by remnants of the Luttinger liquid. In fact, at $k_{\bot }%
\negthinspace =\negthinspace \pi /2$ there is not even a quasiparticle in
this model. 
\begin{figure}
\centerline{\epsfxsize 6cm \epsffile{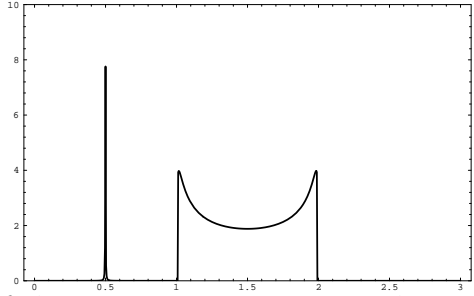}}
\caption{Qualitative sketch of the zero-temperature spectral weight at fixed $k$, as function of frequency 
in the presence of $t_{\bot}$ 
for the simple model with only $g_4$ different from zero and $k_{\bot} < {\pi \over 2}$} 
\label{PicIncoherent}

\end{figure}
In the more general case $\theta \negthinspace \neq \negthinspace 0$, we
cannot repeat the above derivation to find $Z\left( t_{\bot }\right) $
analytically although we can prove from known $1D$ spectral weights \cite
{Voit} that the same reasoning would predict a pole in regions where the $1D$
spectral weight is zero, leading to results qualitatively similar to those
just discussed. {\it (End simple model)}
\end{description}

To find the temperature scale $T_{x^1}$ at which the pole in ${\cal G}^{2D}$
becomes perceptible and transverse single-fermion coherence starts to
develop, we use the natural change of variables $\tau ^{\prime }=\pi T\tau $
and $x=(\xi _\rho \xi _\sigma )^{1/2}x^{\prime }$ to evaluate the Fourier
transform of the $1D$ Green's function (\ref{luttpropa}) at the Fermi level $%
G^{(1)}(0,\pi T)$. Substituting in the $2D$ Gaussian propagator Eq.(\ref{six}%
) one readily finds 
\begin{equation}
\label{Tx1}T_{x^1}\sim E_F\left( {\frac{t_{\bot }}{E_F}}\right) ^{1/\left(
1-\theta \right) }\left( {\frac{v_F}{v_\rho }}\right) ^{\theta _\rho /\left(
1-\theta \right) }\left( {\frac{v_F}{v_\sigma }}\right) ^{\theta _\sigma
/\left( 1-\theta \right) }F_1\left( \left\{ {v_\sigma /v_\rho }\right\}
^{1/2}\right) 
\end{equation}
where $F_1\left( x\right) $ is a temperature independent function that
satisfies $F_1\left( x\right) =F_1\left( 1/x\right) $ and which also depends
on $\theta _{\rho ,\sigma }$. As long as $\theta \negthinspace
<\negthinspace 1$, or equivalently if $G^{(1)}(x)$ decays more slowly than $%
x^{-2}$, the coupling $t_{\bot }$ is relevant and $T_{x^1}$ is finite,
although smaller than the non-interacting value $T_{x^1}\sim t_{\bot }/\pi $.%
\cite{Bourbon1,Bourbon2,Castel92} The condition $\theta \negthinspace <%
\negthinspace 1$ is satisfied for the Hubbard model with a non half-filled
band where one has the exact result $\theta \negthinspace \le \negthinspace {%
1/8}$ \cite{Schulz}. For more specialized 1D models (forward scattering
only, half-filling, etc.) one can have the cases $\theta \negthinspace =%
\negthinspace 1$ and $\theta \negthinspace
>\negthinspace 1$ where transverse hopping becomes marginal and irrelevant
respectively. In these cases, transverse band motion does not develop and
the electrons remain spatially confined along the chains at all
temperatures. As seen from Eq.(\ref{Tx1}), the effect of $\Delta v\neq 0$ is
to decrease the deconfinement temperature but not to make it vanish. The
vanishing of $T_{x^1}$ is nevertheless expected for sufficiently strong
coupling since spin and charge degrees of freedom must recombine for an
electron to tunnel on a neighboring chain. Indeed, it can be shown that in
the limit where the ratio of spin to charge velocities vanish, then the
dimensionless function behaves as $F_1\left( \left\{ v_\sigma /v_\rho
\right\} ^{1/2}\right) \stackrel{v_\sigma /v_\rho \rightarrow 0}{\sim }%
\left\{ v_\sigma /v_\rho \right\} ^{1/2}$ leading in turn to $%
T_{x^1}\rightarrow 0$ when the condition $\theta _\sigma <\frac 13(1-\theta
_\rho )$ is also satisfied.

\subsubsection{In a magnetic field.}

Dimensional crossover in the presence of a magnetic field has been the
subject of numerous studies recently\cite{Hubert}. To illustrate the wide
applicability of our method, we just comment, in a few special cases, on the
effect of a magnetic field on the one-particle dimensional crossover.
Working in the Landau gauge, and using the Peierls substitution 
\begin{equation}
t_{\bot }\rightarrow t_{\bot }\exp \left( \frac{ie}c\int {\bf A\cdot }d{\bf r%
}\right) \rightarrow t_{\bot }e^{iQx} 
\end{equation}
with $Q=ed_{\bot }B/c$, the condition for the dimensional crossover becomes 
\begin{equation}
t_{\bot }\left| G_p\left( k_F-Q,i\pi T\right) \right| \sim 1 
\end{equation}
In the case of non-interacting electrons, this gives that the one-particle
crossover temperature obeys 
\begin{equation}
\left( \pi T_{x^1}^0\right) ^2+\left( v_FQ\right) ^2\sim t_{\bot }^2 
\end{equation}
For large enough magnetic field then ($Q_c^0=ed_{\bot }B_c/c=t_{\bot }/v_F$%
), the one-particle dimensional crossover does not occur. The electrons
remain one-dimensional.

In the simple $g_4$ model, the magnetic field that leads to confinement is
smaller. 
\begin{equation}
Q_c=ed_{\bot }B_c/c=\frac{t_{\bot }}{\left( v_\rho v_\sigma \right) ^{1/2}}=%
\frac{t_{\bot }/v_F}{\sqrt{1+\frac{g_4}{\pi v_F}}} 
\end{equation}
In other words, the Luttinger liquid in this case is easier to confine than
free electrons.

\subsection{Two-particle dimensional crossover.}

We now proceed beyond the Gaussian level by taking into account the ${\cal O}%
(t_{\bot }^2)$ quartic term in the functional Eq.(\ref{quatre}) which
describes correlated transverse pair tunneling. We argue that the system
will undergo a two-particle dimensional crossover towards CDW, SDW ordered
states if the interaction is repulsive and SS or TS superconducting states
if it is attractive. Focusing on the $2k_F^0$ particle-hole channel,
(repulsive case) we rewrite the partition function at the quartic level as 
\begin{equation}
{Z}=Z_{1D}\langle {e^{\sum\limits_{\mu ,q_{\bot }}\int \{dz\}O_{\mu ,q_{\bot
}}^{*}(z_3,z_1)\gamma _{\bot }(q_{\bot })R_\mu (\{z\})O_{\mu ,q_{\bot
}}(z_2,z_4)}}\rangle _{\psi ^{*}\psi }
\end{equation}
with the obvious notations $\{z\}\negthinspace =\negthinspace
\{z_1,z_2,z_3,z_4\}$ and $\langle \ldots \rangle _{\psi ^{*}\psi }$ as the
average with respect to ${\cal F}_0$. The composite fields ${\it O}_{\mu
,q_{\bot }}(z_3,z_1)=N_{\bot }^{-{1/2}}\sum_{k_{\bot }}\sum_{\alpha \beta
}\psi _{-,k_{\bot }}^{\alpha *}(z_3)\sigma _\mu ^{\alpha \beta }\psi
_{+,k_{\bot }+q_{\bot }}^\beta (z_1)$ describe CDW ($\mu \negthinspace
=\negthinspace 0$) and SDW ($\mu \negthinspace =\negthinspace 1,2,3$)
correlations. For nearest-neighbor hopping, we approximate the transverse
pair tunneling amplitude as $\gamma _{\bot }(q_{\bot })\approx (2t_{\bot
})^2\cos (q_{\bot }),$ where $q_{\bot }$ is the transverse momentum of the
particle-hole pair, by setting the incoming momenta to $0$ or $\pi $ since
this leads to the highest value for $T_{x^2}$. In the above, $\sigma
_0^{\alpha \beta }=\delta ^{\alpha \beta }$, $\sigma _{\mu =1,2,3}^{\alpha
\beta }$ are the Pauli matrices and the $q_{\bot }$ independent function $%
R_\mu (\{z\})$ is the appropriate linear combination of connected four point
functions corresponding to charge or spin fluctuations.

To examine the possibility of phase transition, we perform a
Hubbard-Stratonovich transformation on the ${\it O}_\mu $ fields. Let $\zeta
_{\mu ,q_{\bot }}^{*}(z_3,z_1)$ be the complex field conjugate to ${\it O}%
_{\mu ,q_{\bot }}(z_3,z_1)$. The partition function $Z$ then takes the form%
\begin{eqnarray}\label{espoir}
Z=Z_{1D}\int\!\!\!\!\int{\cal D}\zeta^*{\cal D}\zeta\,
e^{-\;\int\! \{dz\}\! \sum\limits_{\mu,q_{\bot}}
\zeta_{\mu,q_{\bot}}^{*}(z_3, z_1) 
\bigl(\openone-\gamma_{\bot}(q_{\bot}){\mit R}_{\mu}(\{z\})\bigr)
\zeta_{\mu,q_{\bot}}(z_2, z_4)+{\cal O}(\zeta^4)}.
\end{eqnarray}
Softening of the $\zeta $ field first occurs for a value of $q_{\bot }%
\negthinspace =\negthinspace \pi $ corresponding to the usual staggered
order. The temperature $T_{x^2}$ at which the $\zeta $ field softens must be
greater than $T_{x^1}$ to retain its meaning as a two-particle crossover
temperature. In a very rough manner of speaking, the Hubbard-Stratonovich
transformation allows us write, by analogy with the one-particle case Eq.(%
\ref{six}) 
\begin{equation}
{\cal G}_{2D}^{\left( 2\right) }\sim {\frac{{G^{(2)}}}{1-t_{\bot }^2G_c^{(2)}%
}} 
\end{equation}
so that the two-particle crossover temperature occurs at $t_{\bot
}^2G_c^{(2)}\sim 1.$ Let us be more precise.

In the weak-coupling regime, $2-2\theta -\gamma _\mu >0$, the correlator
essentially scales as the square of the one-particle propagator (\ref
{luttpropa}) so that 
\begin{equation}
\label{Tx2b}T_{x^2}^\mu \sim T_{x^1}F_3^\mu \left( \left( v_\sigma /v_\rho
\right) ^{1/2},\left\{ g\right\} \right) . 
\end{equation}
where $\left\{ g\right\} $ stands for the set of dimensionless
electron-electron interaction vertices and $F_3^\mu (x,y)$ vanishes when
either of its arguments vanishes. It is only for sufficiently weak coupling
that $T_{x^2}<T_{x^1}$ and that deconfinement takes place.

In the gapless strong-coupling limit $2-2\theta -\gamma _\mu <0$ , we
already know that the correlator $R_\mu (z_1-z_3,z_2-z_4,z_1-z_2)$ decays
faster than the square of the electron-hole separations $|z_1-z_3|$ and $%
|z_2-z_4|$ so that the scaling is determined by $z=z_1-z_2$ only, leading to
the asymptotic form in Eq.(\ref{repos}). To show that the fast decay of $%
R_\mu (z_1-z_3,z_2-z_4,z_1-z_2)$ leads effectively to a contraction of the
two coordinates appearing in the Hubbard-Stratonovich field requires a
discussion whose details appear in Ref.(\cite{TheseBoies}). We find that in
this regime the two-particle crossover temperature is given by 
\begin{equation}
\label{Tx2a}T_{x^2}^\mu \sim E_F\left( {\frac{t_{\bot }^2}{E_F^2}}\right)
^{1/\gamma _\mu }\left( {\frac{v_F}{v_\rho }}\right) ^{\gamma _{\mu ,\rho
}/\gamma _\mu }\left( {\frac{v_F}{v_\sigma }}\right) ^{\gamma _{\mu ,\sigma
}/\gamma _\mu }F_2^\mu \left( \left( v_\sigma /v_\rho \right) ^{1/2}\right)
\;
\end{equation}
where $E_F\sim $ $\Lambda v_F$ while $\gamma _\mu $ is the susceptibility
exponent {$R_\mu(2k_F^0,T)\sim T^{-\gamma _\mu }.$} The 
dimensionless function $F_2^\mu (x)$ vanishes at $x=0$.

The third and last regime is the strong-coupling regime with a gap. In the
case of a charge gap, the asymptotic expression for the correlator is given
by Eq.(\ref{gaprho}), from which we find 
\begin{equation}
\label{Tx2c}T_{x^2}^\mu \sim E_F\left( {\frac{t_{\bot }^2}{E_F^2}}\right) ^{%
\frac 1{1-2\theta _\sigma }}\left( \frac{E_F}{\Delta _\rho }\right) ^{\frac{%
1-2\theta _\rho }{1-2\theta _\sigma }}\left( {\frac{v_F}{v_\rho }}\right) ^{%
\frac{2+2\theta _\rho }{1-2\theta _\sigma }}\left( {\frac{v_F}{v_\sigma }}%
\right) ^{\frac{2+2\theta _\sigma }{1-2\theta _\sigma }}F_5^\mu \left(
\theta _\sigma \right) \; 
\end{equation}
Clearly, the increase of the gap $\Delta _\rho $ with increasing coupling
makes $T_{x^2}^\mu $ decrease with increasing coupling, as stated in the
introduction, unless $\left( 1-2\theta _\rho \right) /\left( 1-2\theta
_\sigma \right) <0$

The above expressions for the two-particle crossover temperature Eqs.(\ref
{Tx2a})(\ref{Tx2b}) and (\ref{Tx2c}) reduce to the known results \cite
{Bourbon1}\cite{Bourbon2} when $v_\rho \negthinspace =\negthinspace v_\sigma 
$. 

Higher-order vertices, ${\cal O}(t_{\bot }^3)...$ involve higher-order
connected functions $G_c^{(3)}...$ of the Luttinger liquid. As mentioned in
the section on the Luttinger liquids, these functions do not contain any new
relevant exponent, hence they do not change the above scaling results for
either $T_{x^1}$ or $T_{x^2}$. In the latter case, higher-order connected
functions modify mode-mode coupling terms included in the functional (\ref
{espoir}).

\section{Comparison with experiment.}

The quasi one-dimensional compound $(TMTTF)_2Br$ at ambient pressure has a
resistivity that increases rapidly below a resistivity minimum at a
temperature $T_\rho $. This signals the opening of a charge gap. Klemme {\it %
et al.}\cite{Klemme} have measured this temperature $T_\rho $ as a function
of pressure (diamonds on Fig.(\ref{Experience}) below). The results are
presented by J\'erome at this conference. Using NMR, they have also measured
the N\'eel temperature for magnetic order $T_N$ (squares) as a function of
pressure. One should understand that increasing pressure increases the
overlap between orbitals, hence it decreases the dimensionless coupling
constants. This means that when one compares the data in Fig.(\ref
{Experience}) with the schematic representation of our results Fig.\ref
{Diagramme}, one should move from right to left in the experimental plot
when the interaction increases on the theoretical plot. The qualitative
agreement between theory and experiment is excellent. In particular, it is
clear from the experiment that the decrease of $T_N$ with increasing
interaction (decreasing pressure) starts only when the charge gap opens up.
At higher pressures, when there is no resistivity minimum above the N\'eel
temperature, $T_N$ always increases with decreasing pressure. The inset on
the figure emphasizes the low temperature part of the diagram to make the
maximum in N\'eel temperature more apparent.%
\begin{figure}
\centerline{\epsfxsize 6cm \epsffile{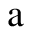}}
\caption{From Klemme et al., Fig.2, Ref.\protect\cite{Klemme}, temperature 
for magnetic order $T_N$ (squares)
as a function of pressure for $(TMTTF)_2Br$. The connection between the maximum in $T_N$ and the 
minimum in resistivity $T_{\rho}$ (diamonds) is apparent.} 
\label{Experience}

\end{figure}

\section{Conclusions.}

Our theoretical conclusions have been summarized at the end of our
introduction. We have further shown that they are consistent with recent
experiments also discussed at this conference.\cite{Klemme} The one-particle
crossover in the presence of a magnetic field has also been briefly
discussed.

We thank M. Fabrizio and J. Voit for helpful discussions. This work was
partially supported by the Natural Sciences and Engineering Research Council
of Canada (NSERC), the Fonds pour la formation de chercheurs et l'aide \`a
la recherche from the Government of Qu\'ebec (FCAR) and (A.-M.S.T.) the
Killam Foundation as well as the Canadian Institute of Advanced Research
(CIAR). A.-M.S.T. would like to thank the Institute for Theoretical Physics,
UCSB. Partial support was provided by the National Science Foundation under
Grant No. PHY94-07194.

\noindent

\begin{center}
LIQUIDES DE LUTTINGER COUPL\'ES.\ \bigskip
\end{center}

\bigskip

La stabilit\'e du liquide de Luttinger en pr\'esence de saut intercha\^\i ne
a \'et\'e \'etudi\'ee de plusieurs points de vue. L'approche bas\'ee sur le
groupe de renormalisation a particuli\`erement \'et\'e critiqu\'ee parce
qu'elle ne tient pas compte explicitement de la diff\'erence entre vitesses
de spin et de charge et aussi parce que si la stabilit\'e du liquide de
Luttinger est un effet non perturbatif alors il faut tenir compte des
interactions avant d'inclure le saut intercha\^\i ne. Une approche qui
r\'epond \`a ces deux objections est d\'ecrite ici. Elle montre que le
liquide de Luttinger est instable en pr\'esence de sauts intercha\^\i nes
arbitrairement petits. Les temp\'eratures de crossover sous lesquelles se
d\'eveloppent soit un mouvement de bande coh\'erent, soit l'ordre \`a longue
port\'ee, peuvent \^etre diff\'erentes de z\'ero m\^eme lorsque les vitesses
de spin et de charge diff\`erent. Des relations d'\'echelle explicites pour
ces temp\'eratures de crossover \`a une-particule ou \`a deux particules
sont d\'eriv\'ees en fonction du saut intercha\^\i ne, des vitesses de spin
et de charge et des exposants anormaux. On retrouve les r\'esultats du
groupe de renormalisation comme cas particulier o\`u les vitesses de spin et
de charge sont identiques. Les r\'esultats se comparent favorablement \`a de
r\'ecentes exp\'eriences pr\'esent\'ees \`a cette conf\'erence. On fait
aussi allusion \`a certains effets des champs magn\'etiques.\bigskip


\begin{references}
\bibitem{Haldane}  F. D. M. Haldane, Phys. Rev. Lett. {\bf 45}, 1358 (1980);
J. Phys. C {\bf 14}, 2585 (1981); Phys. Lett. {\bf 81A}, 153 (1981); Phys.
Rev. Lett. {\bf 47}, 1840 (1981).

\bibitem{Anders}  P. W. Anderson, Phys. Rev. Lett. {\bf 64}, 1839 (1990);
Phys. Rev. Lett. {\bf 65}, 2306 (1990); Phys. Rev. Lett. {\bf 67}, 3844
(1991).

\bibitem{Bourbon1}  C. Bourbonnais {\it et al.,} J. Physique Lett., {\bf 45}%
, L755 (1984); V. N. Prigodin and Y. A. Firsov, Sov. Phys. JETP {\bf 49}%
,369(1979); H. G. Schuster, Phys. Rev. {\bf B 13}, 628 (1976); L.P. Gorkov
and I.E. Dzyaloshinskii, Sov. Phys. JETP {\bf 40}, 198 (1974).

\bibitem{Brazo}  S. Brazovskii and V. Yakovenko, Sov. Phys. JETP {\bf 62},
1340(1985).

\bibitem{Bourbon2}  C. Bourbonnais and L. G. Caron, Physica {\bf B143},
451(1986); Europhys. Lett., {\bf 5}, 209(1988); Int. Journ. Phys. {\bf 5},
1033(1991).

\bibitem{Twochain}  M. Fabrizio, A. Parola and E. Tosatti, Phys. Rev. {\bf B
46}, 3159(1992); M. Fabrizio and A. Parola, Phys. Rev. Lett. {\bf 70},
226(1992); A. Finkelstein and A.I. Larkin, Phys. Rev. {\bf B 47},
10461(1993); V. Yakovenko, JETP lett. {\bf 56}, 5101(1992); A.A. Nerseyan,
A. Luther, and F. V. Kusmartsev, Phys. Lett. {\bf A 176}, 363(1993).

\bibitem{Castel92}  C. Castellani, C. Di Castro, and W. Metzner, Phys. Rev.
Lett. {\bf 69}, 1703 (1992).

\bibitem{Anders3}  D. G. Clarke, S. P. Strong, P.W. Anderson, Phys. Rev.
Lett. {\bf 72}, 3218(1994).

\bibitem{TheseBoies}  PhD Thesis, Daniel Boies, Universit\'e de Sherbrooke
III-920 (1994) (unpublished).

\bibitem{prl}  Daniel Boies, C. Bourbonnais, and A.-M.S. Tremblay, Phys.
Rev. Lett. {\bf 74}, 968 (1995).

\bibitem{These}  C. Bourbonnais, Ph.D thesis, Universit\'e de Sherbrooke,
(1985), unpublished.

\bibitem{LesHouches}  C. Bourbonnais, in {\it Strongly Interacting Fermions
and High }$T_c$ S{\it uperconductivity, }Les Houches, Session LVI, 1991,
Eds. B. Dou\c cot and J. Zinn-Justin (Elsevier, Amsterdam, 1995), p.307.

\bibitem{spinchar}  H. J. Schulz, Int. J. Mod. Phys. B {\bf 5}, 57 (1991).

\bibitem{lieb}  D. C. Mattis et E. H. Lieb, J. Math. Phys. {\bf 6}, 304
(1965).

\bibitem{luther}  A. Luther et J. Peshel, Phys. Rev. B {\bf 9}, 2911 (1974).

\bibitem{unidim} A. M. Finkelshtein, JETP Lett. {\bf 25}, 73 (1977). 

\bibitem{VoitR} For a review, see J. Voit, Reports on Progress in Physics 
{\bf 58}, p. 977 (1995).

\bibitem{geo}  J. Solyom, Adv. Phys. {\bf 28}, 201 (1979).

\bibitem{lutt}  J. M. Luttinger, J. Math. Phys. {\bf 4}, 1154 (1963).

\bibitem{dzyalo}  I. E. Dzyaloshinski et A. I. Larkin, Sov. Phys. JETP {\bf %
38}, 202 (1974).

\bibitem{frahm}  H. Frahm et V. E. Korepin, Phys. Rev. B {\bf 42}, 10553
(1990).

\bibitem{frahmdeux}  H. Frahm et V. E. Korepin Phys. Rev. B {\bf 43}, 5653
(1991).

\bibitem{emery}  V. J. Emery in {\it Highly Conducting One-Dimensional Solids%
}, eds J. T. Devreese, R. P. Evrard et V. E. van Doren (Plenum 1979), p. 247.

\bibitem{casteldeux}  C. Castellani, C. Di Castro, and W. Metzner, Phys.
Rev. Lett.{\bf 72}, 316 (1994).

\bibitem{Voit}  J. Voit, Phys. Rev. {\bf B 47}, 6740(1993);
Habilitationsschrift, Universit\"at Bayreuth, 1993 (unpublished).

\bibitem{brazoun}  S. Brazovskii et V. Yakovenko, J. Physique Lett. (Paris) 
{\bf 46}, p. L-111 (1985).

\bibitem{brazodeux}  S. A. Brazovskii et V. M. Yakovenko, Sov. Phys. JETP 
{\bf 62}, 1340 (1985).

\bibitem{gutklem}  H. Gutfreund and R. A. Klemm, Phys. Rev. B {\bf 14}, 1073
(1976).

\bibitem{Schulz}  H. J. Schulz, Phys. Rev. Lett. {\bf 64}, 2831(1990).

\bibitem{Hubert} L. Hubert and C. Bourbonnais, Synthetic Metals, {\bf 
55-57}, 4231 (1993).

\bibitem{Klemme}  B.J. Klemme, S.E. Brown, P.Wzietek, G. Kriza, P. Batail,
D. J\'erome, J.M. Fabre, Phys. Rev. Lett. {\bf 75}, 2408 (1995).
\end{references}
\end{document}